\documentstyle[12pt]{article}

\catcode`\@=11
\long\def\@makefntext#1{
\protect\noindent \hbox to 3.2pt {\hskip-.9pt  
$^{{\ninerm\@thefnmark}}$\hfil}#1\hfill}		%CAN BE USED 

\def\@makefnmark{\hbox to 0pt{$^{\@thefnmark}$\hss}}  %ORIGINAL 
	
\def\ps@myheadings{\let\@mkboth\@gobbletwo
\def\@oddhead{\hbox{}
\rightmark\hfil\ninerm\thepage}   
\def\@oddfoot{}\def\@evenhead{\ninerm\thepage\hfil
\leftmark\hbox{}}\def\@evenfoot{}
\def\sectionmark##1{}\def\subsectionmark##1{}}

\renewcommand{\thefootnote}{\fnsymbol{footnote}}

\newcounter{sectionc}\newcounter{subsectionc}\newcounter{subsubsectionc}
\renewcommand{\section}[1] {\vspace*{0.6cm}\addtocounter{sectionc}{1} 
\setcounter{subsectionc}{0}\setcounter{subsubsectionc}{0}\noindent 
	{\normalsize\bf\thesectionc. #1}\par\vspace*{0.4cm}}
\renewcommand{\subsection}[1] {\vspace*{0.6cm}\addtocounter{subsectionc}{1} 
	\setcounter{subsubsectionc}{0}\noindent 
	{\normalsize\it\thesectionc.\thesubsectionc. #1}\par\vspace*{0.4cm}}
\renewcommand{\subsubsection}[1]
{\vspace*{0.6cm}\addtocounter{subsubsectionc}{1}
	\noindent {\normalsize\rm\thesectionc.\thesubsectionc.\thesubsubsectionc. 
	#1}\par\vspace*{0.4cm}}

\newcounter{appendixc}
\newcounter{subappendixc}[appendixc]
\newcounter{subsubappendixc}[subappendixc]

\renewcommand{\appendix}[1] {\vspace*{0.6cm}
        \refstepcounter{appendixc}
        \setcounter{figure}{0}
        \setcounter{table}{0}
        \setcounter{equation}{0}
        \renewcommand{\thefigure}{\Alph{appendixc}.\arabic{figure}}
        \renewcommand{\thetable}{\Alph{appendixc}.\arabic{table}}
        \renewcommand{\theappendixc}{\Alph{appendixc}}
        \renewcommand{\theequation}{\Alph{appendixc}.\arabic{equation}}
        \noindent{\bf Appendix \theappendixc #1}\par\vspace*{0.4cm}}

\def\abstracts#1{{
	\centering{\begin{minipage}{12.2truecm}\footnotesize\baselineskip=12pt\noindent
	\centerline{\footnotesize ABSTRACT}\vspace*{0.3cm}
	\parindent=0pt #1
	\end{minipage}}\par}} 

\renewenvironment{thebibliography}[1]
	{\begin{list}{\arabic{enumi}.}
	{\usecounter{enumi}\setlength{\parsep}{0pt}
\setlength{\leftmargin 1.25cm}{\rightmargin 0pt}
	 \setlength{\itemsep}{0pt} \settowidth
	{\labelwidth}{#1.}\sloppy}}{\end{list}}

\newcounter{itemlistc}
\newcounter{romanlistc}
\newcounter{alphlistc}
\newcounter{arabiclistc}

\newcommand{\fcaption}[1]{
        \refstepcounter{figure}
        \setbox\@tempboxa = \hbox{\footnotesize Fig.~\thefigure. #1}
        \ifdim \wd\@tempboxa > 6in
           {\begin{center}
        \parbox{6in}{\footnotesize\baselineskip=12pt Fig.~\thefigure. #1}
            \end{center}}
        \else
             {\begin{center}
             {\footnotesize Fig.~\thefigure. #1}
              \end{center}}
        \fi}

\newcommand{\tcaption}[1]{
        \refstepcounter{table}
        \setbox\@tempboxa = \hbox{\footnotesize Table~\thetable. #1}
        \ifdim \wd\@tempboxa > 6in
           {\begin{center}
        \parbox{6in}{\footnotesize\baselineskip=12pt Table~\thetable. #1}
            \end{center}}
        \else
             {\begin{center}
             {\footnotesize Table~\thetable. #1}
              \end{center}}
        \fi}

\def\@citex[#1]#2{\if@filesw\immediate\write\@auxout
	{\string\citation{#2}}\fi
\def\@citea{}\@cite{\@for\@citeb:=#2\do
	{\@citea\def\@citea{,}\@ifundefined
	{b@\@citeb}{{\bf ?}\@warning
	{Citation `\@citeb' on page \thepage \space undefined}}
	{\csname b@\@citeb\endcsname}}}{#1}}

\newif\if@cghi
\def\cite{\@cghitrue\@ifnextchar [{\@tempswatrue
	\@citex}{\@tempswafalse\@citex[]}}
\def\citelow{\@cghifalse\@ifnextchar [{\@tempswatrue
	\@citex}{\@tempswafalse\@citex[]}}
\def\@cite#1#2{{$\null^{#1}$\if@tempswa\typeout
	{IJCGA warning: optional citation argument 
	ignored: `#2'} \fi}}

 1
 1
 1

\font\ninerm=cmr9

\begin{document}
\begin{titlepage}
%
%footnotesymbols others than numbers
\renewcommand{\thefootnote}{\fnsymbol{footnote}}
\begin{flushright}
%BONN-TH/96-09\\
%Leipzig-Preprint?? \\
Februar 1997
\end{flushright}
\vspace{1cm}
 
\begin{center}
{ {\bf CALLAN-SYMANZIK AND RENORMALIZATION GROUP 
EQUATION IN THEORIES WITH SPONTANEOUSLY BROKEN
SYMMETRY}}\footnote{
To appear in the ``Proceedings of the Ringberg workshop
       1996: The Higgs puzzle -- What can we learn from 
        LEP II, LHC, NLC, and FMC?"; World Scientific Publishing
       Company, ed.~B.~A.~Kniehl.} \\[4mm]
{\makebox[1cm]{  }       \\[2cm]
{ ELISABETH KRAUS}\\ [3mm]
{\small\sl Physikalisches Institut, Universit\"at Bonn} \\
{\small\sl Nu\ss allee 12, D-53115 Bonn, Germany}} 
\vspace{2.0cm}
 
{\bf Abstract}
\end{center}
\begin{quote}
Callan-Symanzik and renormalization group equation 
are discussed for  the $U(1)$-axial model
 and it is shown, that the symmetric model is
not the asymptotic version of the spontaneously broken one due to 
mass logarithms in the $\beta$-functions. The Callan-Symanzik 
equation of the standard
model is seen to have the same form as  the one of the simple model.
\end{quote}
\vfill
\renewcommand{\thefootnote}{\arabic{footnote}}
\setcounter{footnote}{0}
\end{titlepage}

\centerline{\normalsize\bf CALLAN-SYMANZIK AND RENORMALIZATION GROUP 
EQUATION}
\baselineskip=15pt
\centerline{\normalsize\bf IN THEORIES WITH SPONTANEOUSLY BROKEN
SYMMETRY}
\baselineskip=16pt
%\centerline{\normalsize\bf MANUSCRIPT BY COMPUTER}
%\centerline{\footnotesize\sf (For subsequent 20\% photoreduction
%to 17.8 $\times$ 11.9 cm text area) \footnote{The \LaTeX\ source
%file for this document may be used as a template for your
%article, and can be requested by e-mailing {\sf
%worldscp@singnet.com.sg}.}}

%\vfill 
\vspace*{0.6cm}
\centerline{\footnotesize ELISABETH KRAUS\footnote{Supported by
Deutsche Forschungsgemeinschaft}}
\baselineskip=13pt
\centerline{\footnotesize\it Universit\"at Bonn, Nu\ss allee 12
}
\baselineskip=12pt
\centerline{\footnotesize\it D-53115 Bonn, Germany}
\centerline{\footnotesize E-mail: kraus@avzw02.physik.uni-bonn.de}
\vspace*{0.3cm}
%\centerline{\footnotesize and}
%\vspace*{0.3cm}
%\centerline{\footnotesize SECOND AUTHOR'S NAME}
%\baselineskip=13pt
%\centerline{\footnotesize\it Group, Company, Address, City, State ZIP/Zone,
%Country}

%\vfill
\vspace*{0.9cm}
\abstracts{CS and RG equation are discussed for  the $U(1)$-axial model
 and it is shown, that the symmetric model is
not the asymptotic version of the spontaneously broken one due to 
mass logarithms in the $\beta$-functions. The CS-equation of the standard
model is seen to have the same form as  the one of the simple model.} 
 
%\vspace*{0.6cm}
\normalsize\baselineskip=15pt
\setcounter{footnote}{0}
\renewcommand{\thefootnote}{\alph{footnote}}
\newcommand{\N}{{\cal N}}
\newcommand{\Cal}{{\cal C}}
\renewcommand{\l}{\lambda}
\renewcommand{\a}{\alpha}
\renewcommand{\b}{\beta}
\renewcommand{\d}{\delta}
\renewcommand{\k}{\kappa}
\newcommand{\ld}{\buildrel \leftarrow \over \d}
\newcommand{\rd}{\buildrel \rightarrow \over \d}
\newcommand{\e}{\eta}
\renewcommand{\o}{\omega}
\newcommand{\dem}{\d_\o}
\newcommand{\p}{\partial}
\newcommand{\pmu}{\p_\mu}
\newcommand{\pmo}{\p^\mu}
\newcommand{\pnu}{\p_\nu}
\newcommand{\s}{\sigma}
\renewcommand{\r}{\rho}
\newcommand{\bpsi}{\bar\psi}
\newcommand{\dslash}{\p\llap{/}}
\newcommand{\pslash}{p\llap{/}}
\newcommand{\ve}{\varepsilon}
\newcommand{\uvi}{\underline{\varphi}}
\newcommand{\vi}{\varphi}
\newcommand{\ue}{u^{(1)}}
\newcommand{\Am}{A_\mu}
\newcommand{\An}{A_\nu}
\newcommand{\Fmnu}{F_{\mu\nu}}
\newcommand{\Fmno}{F^{\mu\nu}}
\newcommand{\Ga}{\Gamma}
\newcommand{\Gao}{\Gamma^{(o)}}
\newcommand{\Gae}{\Gamma^{(1)}}
\newcommand{\Gacl}{\Gamma_{cl}}
\newcommand{\Gagf}{\Gamma_{{\rm g.f.}}}
\newcommand{\Gainv}{\Gamma_{\rm inv}}
\newcommand{\pvi}{\partial\varphi}
\newcommand{\om}{{\bf w}}
\newcommand{\mn}{\mu\nu}
\newcommand{\Tmn}{T_{\mn}}
\newcommand{\Gmn}{\Ga_{{\mn}}}
\newcommand{\hT}{\hat T}
\newcommand{\emt}{energy-mo\-men\-tum ten\-sor}
\newcommand{\eit}{Ener\-gie-Im\-puls-Ten\-sor}
\newcommand{\Tc}{T^{(c)}}
\newcommand{\Tcr}{\Tc_{\rho\sigma}(y)}
\newcommand{\ha}{{1\over 2}}
\newcommand{\dalam}{{\hbox{\frame{6pt}{6pt}{0pt}}\,}}
\newcommand{\wtm}{{\bf W}^T_\mu}
\newcommand{\nnp}{\nu'\nu'}
\newcommand{\mnp}{\mu'\nu'}
\newcommand{\emn}{\eta_{\mn}}
\newcommand{\mpm}{\mu'\mu'}
\newcommand{\tbw}{\tilde{\bf w}}
\newcommand{\tw}{\tilde w}
\newcommand{\hw}{\hat{\bf w}}
\newcommand{\bw}{{\bf w}}
\newcommand{\bW}{{\bf W}}
\newcommand{\ubW}{\underline{\bW}}
\newcommand{\hmn}{h^{\mn}}
\newcommand{\gmn}{g^{\mn}}
\newcommand{\ga}{\gamma}
\newcommand{\Gf}{\Gamma_{\hbox{\hskip-2pt{\it eff}\hskip2pt}}}
\newcommand{\T}{\buildrel o \over T}
\newcommand{\Lf}{{\cal L}_{\hbox{\it eff}\hskip2pt}}
\newcommand{\np}{\not\!\p}
\newcommand{\lp}{\partial\llap{/}}
\newcommand{\ah}{{\hat a}}
\newcommand{\han}{{\hat a}^{(n)}}
\newcommand{\hak}{{\hat a}^{(k)}}
\newcommand{\dv}{{\d\over\d\vi}}
\newcommand{\zze}{\sqrt{{z_2\over z_1}}}
\newcommand{\zez}{\sqrt{{z_1\over z_2}}}
\newcommand{\Hmn}{H_{\(\mn\)}}
\newcommand{\smdm }{\underline m \p _{\underline m}}
 \newcommand{\tsmdm }{\underline m \tilde \p _{\underline m}} 
 \newcommand{ \Wh }{{\hat {\bf W}}^K}
\newcommand{ \CS }{Callan-Symanzik}
\newcommand{ \G}{\Gamma}
\newcommand{ \bl }{\b _ \l}
\newcommand{ \kdk }{\k \p _\k}
\newcommand{\mdm }{m \p _m} 
\newcommand{ \te }{\tau_{\scriptscriptstyle 1}}
\newcommand{ \mhi }{m_H}
\newcommand{ \mf }{m_f}
\newcommand{\brs}{\mathrm s}
\newcommand{\cw}{\cos \theta_W}
\newcommand{\cws}{\cos^2 \theta_W}
\newcommand{\sw}{\sin \theta_W}
\newcommand{\sws}{\sin^2 \theta_W}
\newcommand{\cg}{\cos \theta_G}
\newcommand{\sg}{\sin \theta_G}
\newcommand{\cwg}{\cos (\theta_W- \theta_G)}
\newcommand{\swg}{\sin (\theta_W- \theta_G)}
\newcommand{\cv}{\cos \theta_V}
\newcommand{\sv}{\sin \theta_V}
\newcommand{\cvg}{\cos (\theta_V- \theta_G)}
\newcommand{\svg}{\sin (\theta_V- \theta_G)}
\newcommand{\fsc}{{e^2 \over 16 \pi ^2}}

\section{Introduction}
All perturbative calculations
suffer from breaking off at a finite, even low, order of loops, whereas the
perturbative expansion is an infinite power series in the couplings.
In order to be able to sum up higher order large contributions, one
can use renormalization group (RG) invariance and the Callan--Symanzik 
(CS) equation, which have been successfully applied in symmetric models
as it is QED.
However, if the theory is spontaneously broken, RG invariance
 can be formulated only for the coupling, which is
the perturbative expansion parameter. But the CS equation, which
describes the breaking of dilatations, involves all interactions which
are differently renormalized and contains therefore the $\beta$-functions
of mass parameters.  We will show, that
in the spontaneously broken model, the $\beta$-functions depend 
logarithmically on the mass parameters, also in the asymptotic region,
from 2-loop order onwards. The symmetric massless model is therefore
not reached in the asymptotic limit. In the last section we
give the CS equation of the standard model and the 
1-loop $\beta$-functions of the  gauge interactions in the on-shell
scheme. It is seen, 
that it involves mass $\beta$-functions in 1-loop order.  Therefore an
 analysis of the symmetric
massless $SU(2)\times U(1)$ theory is not sufficient for determining leading
contributions of higher orders. 

\section{ The
spontaneously broken $U(1)$-axial model}
The $U(1)$-axial model is the simplest model with two different masses,
which are generated by the spontaneous symmetry breaking.
It can be considered as a toy
  model of the matter sector of the standard model, when the
gauge interactions are turned off.
The model describes the interaction of a 
massive scalar $H$, a massless pseudoscalar
$\chi$ and one massive fermion $\psi$.
 The classical
action of the model   is given by:
 \begin{eqnarray}
\label{F6}
\Gacl & = & \int
\Bigl(  \ha \bigl(\p H \p H + \p \chi \p \chi \bigr) +  i \bpsi \dslash  \psi
- m_f \bpsi \psi - {m_f
\over m_  H}
\sqrt {\frac { \l}{3}  } \bpsi (H +  i \gamma _ 5 \chi )\psi \nonumber\\
  & &  - \frac 12  m_  H^2 H^2 - \frac 12
  m_  H \sqrt{ \frac \l 3 } H ( H^2 + \chi^ 2) -
 \frac  \l{4!}    ( H^2 + \chi^ 2)^2
\Bigr)
\end{eqnarray}
As indicated by the notation we will consider the model throughout in the
on-shell parametrization, where the masses of the particles are fixed at the
poles of the respective propagators. 
%There is left   one coupling, which
%is the  expansion parameter of the perturbative calculations.
The invariance of the
classical action under the axial $U(1)$-transformations can be expressed
in terms of a Ward identity:
\begin{equation}
\label{F2}
 \bW \G = 0 \quad \hbox{with} \quad
\bW = -i \int \bigl( ( H + v )  {\d \over \d \chi} - \chi {\d \over \d H} -
                       {i \over 2 }  {\d \over \d \psi }    \ga _5 \psi
                   - {i \over 2 }\bpsi  \ga _5 {\d \over \d \bpsi }
\bigr)
\end{equation}
The Ward identity  (\ref{F2}) together with appropriate normalization
conditions uniquely determine the Green functions to all orders of
perturbation theory.
\begin{eqnarray}
\label{norm}
\p _{p^2} \Gamma_{HH} \big|_ {p^2 = \kappa ^2} = 1
 & \qquad & \gamma^ \mu \p _{ p ^\mu} \Gamma _ {\psi \bpsi}
 \big| _ { \pslash = \kappa} = 1
 \qquad
\Gamma_ H = 0 \nonumber \\
 \Gamma _{HH} \big| _{p^2 = m_H^2} = 0
& \qquad & \Gamma _ {\bpsi \psi} \big|_
{\pslash = m_f} = 0 \qquad
 \G _{HHHH}\big|_ {p^2  =  \kappa^2 } 
= -  \lambda  
\end{eqnarray}
 We have fixed the residua and the coupling at an Euclidean
normalization point $\kappa^2$. The 4-point function of the
Higgs interaction is evaluated at a symmetric momentum $p^2$
defined by
\begin{equation}
\label{sym}
p_i^2 = p^2\, , \qquad p_i\cdot p_j = - \hbox{$\frac 23$} p^2 \quad
i \neq j\,.
\end{equation}
 With these normalization conditions the Green functions are calculated
perturbatively in powers of the coupling $\sqrt \l$, the shift parameter is
determined by the Ward identity:
\begin{equation}
\label{shift}
 v \equiv  v (m_H,m_f,\k
, \l ) = \sqrt { \frac {  3}{ \l} } m_H + O (\hbar )
\end{equation}
The  CS equation describes the breaking of the dilatations 
for the off-shell Green functions in form of a differential 
equation\cite{CaSy}.
In the tree approximation dilatations are broken by the mass terms, this
breaking is related to the differentiation with respect to the
Higgs due to the spontaneous symmetry mechanism and is determined
by the covariance with respect to the Ward identity (\ref{F2}):
\begin{equation}
\label{tree}
m\p _m \Gacl=  \int  \Bigl( v {\delta \Gacl \over \delta H } + {m_H^2 
\over 2}
(H^2 + \chi^2 +
2 v H ) \Bigr) 
\end{equation}
 $m\p _m $ is the dilatational operator when acting on the Green
functions 
\begin{equation}
\label{dilop}
m \p_m  \equiv
  m_{H} \p _{m_{H}}
+  m_{f} \p _{m_{f}}
+\k \p _\k 
\end{equation}

In higher orders dilatations are broken
by anomalies which are given by the $\beta$-functions and 
anomalous dimensions. The CS equation is derived to all orders
 of perturbation theory\cite{CalSym}:
\begin{equation}
\label{G46}
   \bigl(  m \p_m  +  
 \b_{m_f} m_f
 \p _ {m_{f}}  + \b _\l \p _\l -
\ga _ S \N _ S - \ga_F \N _ F
\bigr) \G =
 v (1 + \rho ) \int \bigl( {\d \over {\d H}}
 +\hat \a  {\d \over {\d q}}\bigr) \G_{\big|_{q=0}}
\end{equation}
$\N_S $ and $\N_F $ are the leg counting operators of the scalars and
fermions respectively. The r.h.s.~is the unique continuation of (\ref{tree})
to higher orders, where 
 $q$ is an external field coupled to the invariant of UV dimension 2.

%\begin{eqnarray}
%  \N _S & = & \int \bigl( H{ \d \over \d H} + \chi { \d \over \d \chi}\bigr)
% \qquad
% \N _F =
% \int \bigl( \psi{\d \over \d \psi } + \bpsi {\d \over \d \bpsi} \bigr)
% \nonumber \\
%   v \rho  & = &
%( \b _ \l \p_\l  +   \b_{\mf} \mf \p _ {\mf} + \ga_B)  v \quad
%\hat \alpha = {\mhi^2 \over 2} + O(\hbar)
%\end{eqnarray}
The $\beta$-function with respect to the mass parameter is special for
models with spontaneous breaking of the symmetry. Due to the fact that
the mass differentiation
 acts not only on the vertices but also on propagators it brings about,
that mass logarithms appear in the $\beta$-functions of the CS equation
from 2-loop order onwards. Finally
we want to mention that the r.h.s.~of the CS equation potentially contains 
non-integrable infrared divergencies, which have to be shown to be absent
by an explicit test on  the 2-point function of the massless
particles. 

The RG equation 
describes the response to a change of the arbitrary normalization
point $\kappa$. Due to the physical normalization
conditions it has the simple form
\begin{equation}
\label{G48}
\Bigl( \kappa \p _ \kappa + \tilde  \b _ \l \p _ \l -\tilde \ga _ S
\N _ S  - \tilde  \ga _ F \N_ F \Bigr) \G = 0
\end{equation}
 with the additional constraint:
\begin{equation}
\label{G49}
 \kappa \p _ \kappa  v + \tilde \b_ \l \p _\l  v  +   \tilde
 \ga _ S  v
= 0
\end{equation}
 One has to note that the CS and RG equation do not contain the
same differential operators and that the $\beta$-functions and 
anomalous dimensions are not equal in general. In order to get
insight into mass parameter dependence in higher orders they have
to be solved consistently at the same time. For this purpose we
want to consider in the next section the consequences of RG invariance.

\section{Renormalization group invariance}
In order to simplify the analysis we want to restrict the considerations
of RG invariance to the invariant charge, which is defined in the
$U(1)$-axial model by the following combinations of Green functions
evaluated at the symmetric point $p^2$ (\ref{sym}):
\begin{equation}
\label{qdef}
 Q(\hbox{$  p^2 \over \kappa^2 $},\hbox{$ m^2 \over p^2 $}, \l)  =  
  - \Ga_4 (p^2,m^2,\k^2,\l)
 \biggl( \p _{p^2} 
\Ga_2 (p^2,m^2,\k^2, \l)
\biggr)^{-2}
\end{equation}
 It satisfies the homogeneous RG equation and is therefore a RG invariant
\begin{equation}
\label{rgq}
\bigl(  \kappa \p _ \kappa +\tilde \b _\l \p _\l  
\bigr) Q(\hbox{${p^2\over \k^2}$},\hbox{${m^2 \over p^2}$} , \l) = 0 \, .
\end{equation}
According to (\ref{norm}) the invariant charge
 has well-defined normalization properties 
\begin{equation}
\label{normq}
 Q(\hbox{$  p^2 \over\kappa^2 $},\hbox{$ m^2 \over p^2 $}, \l)  
 \Big| _{p^2 = \k^2 } = \l .
\end{equation}

Finite RG transformation of the invariant charge
can be derived by a formal integration of the RG equation\cite{LowRG},
 but on a
much more fundamental level, invariance of the Green functions under
finite RG transformations up to field redefinitions can be postulated
directly\cite{Bogo,ZiRG}: 
If one fixes the invariant charge at a different 
normalization point ${\kappa'}^2$ to a different coupling
$\lambda '$
\begin{equation}
\label{ncq} 
 Q(\hbox{${p^2\over {\k'}^2}$},\hbox{${m^2 \over p^2}$} , \l' 
)\Big|_{p^2={\k'}^2}
 = \l'
\end{equation}
 RG invariance requires that for all momenta the result has to be the same:
\begin{equation}
\label{rgdef}
 Q(\hbox{${p^2\over \k^2}$},\hbox{${m^2 \over p^2}$} , \l)= 
 Q(\hbox{${p^2\over {\k'}^2}$},\hbox{${m^2 \over p^2}$} , \l')
\end{equation}
Therefrom one derives the multiplication law of the RG
\begin{equation}
\label{rgass}
Q\bigl(\frac{p^2}{\kappa^2} , \frac {m^2}{p^2}
, \l\bigr) = Q\bigl(\frac {p^2}{\kappa'^2}, \frac {m^2}{p^2}, Q\bigl(\frac 
{\kappa'^2}{\kappa^2} , 
\frac {m^2}{\kappa'^2}  , \l\bigr)\bigr)
\end{equation}
%where
%\begin{equation}
%\tau = \frac {p^2}{{\k'}
%^2} , \quad \tau' = \frac {{\k '}^2}{\k^2} , \quad
% u =  \frac {m^2}{p^2}.
%\end{equation}
 Restricting to perturbation theory, where the
invariant charge is calculated in powers of the coupling $\l$,
\begin{equation}
Q(\hbox{$ { p^2 \over \kappa^2} $},\hbox{$ {m^2 \over p^2} $}
, \l)  =  
 \sum_{i=0}^\infty \l^{i+1} f_i(\hbox{$ {p^2 \over \k^2} $},
\hbox{$ {m^2 \over p^2} $} ) \, ,
\end{equation}
one is able to solve the multiplication law of RG order by order
in perturbation theory\cite{InvCh}:
\begin{eqnarray}
\label{rginv}
f_1(\hbox{$ {p^2 \over \k^2} $} ,\hbox{$ {m^2 \over p^2} $}) & =& 
 g_1(\hbox{$ {m^2 \over \k^2} $} ) - g_1 (
\hbox{$ {m^2 \over p^2} $}) \nonumber \\
f_2(\hbox{$ {p^2 \over \k^2} $} ,\hbox{$ {m^2 \over p^2} $}) & =& g_2 (\hbox{$ {m^2 \over \k^2} $}) - g_2(
\hbox{$ {m^2 \over p^2} $}) + (g_1(\hbox{$ {m^2 \over \k^2} $} ) - g_1 (
\hbox{$ {m^2 \over p^2} $}
))^2 
\nonumber \\
f_3(\hbox{$ {p^2 \over \k^2} $},\hbox{$ {m^2 \over p^2} $})  & = 
& g_3 (\hbox{$ {m^2 \over \k^2} $}) - g_3 (\hbox{$ {m^2 \over p^2} $}) 
          +  \frac  52 (g_2 (\hbox{$ {m^2 \over \k^2} $}) -  g_2(
\hbox{$ {m^2 \over p^2} $}))( g_1 (\hbox{$ {m^2 \over \k^2} $}) - g_1(
\hbox{$ {m^2 \over p^2} $})) \nonumber \\ & & + 
              \frac 12  (g_2 (\hbox{$ {m^2 \over \k^2} $})  g_1 ( 
\hbox{$ {m^2 \over p^2} $}) - g_2 ( \hbox{$ {m^2 \over p^2} $}) g_1 (
\hbox{$ {m^2 \over \k^2} $}))
          +  ( g_1 (\hbox{$ {m^2 \over \k^2} $}) - g_1(\hbox{$ {
m^2 \over p^2} $}))^3 
\end{eqnarray}
Therefrom it is seen that RG invariance structures the invariant charge
according to its mass and momentum dependence. If the  functions
$g_i (u) $  are not constant lower order functions are induced to
higher orders and
  RG invariance can only be realized to all
orders of perturbation theory.  

From RG invariance it is not possible to gain information about the
functions $g_i(u)$. They have to be calculated in
perturbation theory by the respective loop diagrams. However, because
one has derived the CS equation, it is possible to find restrictions
on the high energy behavior of these functions. Considering theories
without spontaneous symmetry breaking, as it is $\phi^4$ and QED, the
CS equation involves the same differential operators as the RG equation
(\ref{rgq}):
\begin{equation}
\label{csq}
\bigl(  m \p _ m + \b _\l \p _\l  
\bigr) Q(\hbox{${p^2\over \k^2}$},\hbox{${m^2 \over p^2}$} , \l) = 
Q_m(\hbox{${p^2\over \k^2}$},\hbox{${m^2 \over p^2}$} , \l)  
\end{equation}
\pagebreak[2]
 $Q_m(\hbox{${p^2\over \k^2}$},\hbox{${m^2 \over p^2}$} , \l)  $ denotes
a soft insertion, which vanishes for asymptotic non-exceptional momenta.
Applying the CS equation to the invariant charge as it is given
by RG invariance (\ref{rginv}) one finds that all the functions
$g_i(u)$ tend to logarithms in the asymptotic region:
\begin{equation}
 g_i(\hbox{$m^2 \over p^2$}) {\longrightarrow} \frac 12 b_i 
\ln|\hbox{$m^2 \over p^2$}| \quad \hbox{for} \quad p^2 \to -\infty 
\end{equation}
Moreover one can show that for an asymptotic normalization point
$\kappa^2$ the massless limit is reached smoothly, especially
the $\beta$-functions of the RG and CS equation are equal to
all orders of perturbation theory and mass independent\cite{KrAs}:
\begin{equation}
\beta_\lambda = \tilde \beta_\lambda \quad \hbox{for} \quad \k^2 \to -\infty 
\end{equation}
These are the intrinsic normalization conditions of the
MS- and $\overline{\hbox{MS}}$-scheme.

\section{RG invariance and CS equation in the $U(1)$-axial model}
In order to get insight into the specific structure of the spontaneously
broken models we consider the invariant charge of the Higgs coupling
as it is defined in (\ref{qdef}). It satisfies the homogeneous
RG equation (\ref{rgq}) and for an asymptotic symmetric momentum the
CS equation (\ref{G46})
\begin{equation}
\label{G46i}
   \bigl(  m \p_m  +  
 \b_{m_f} m_f
 \p _ {m_{f}}  + \b _\l \p _\l 
\bigr) Q(\hbox{${p^2\over \k^2}$},\hbox{${m^2 \over p^2}$} ,
 \l)   \longrightarrow 
 0
\quad \hbox{for} \quad p^2 \to -\infty  \,.
\end{equation} 
As a further simplification we take the normalization point, which
normalizes the coupling and the residua, also in the asymptotic 
region. Such a choice corresponds to an on-shell definition for
the masses and a MS definition for residua and the coupling.
It can be shown that for asymptotic momenta the $\beta$-functions
of the coupling in the RG and CS equation are equal
\begin{equation}
\beta_\lambda = \tilde \beta_\lambda \quad \hbox{for} \quad \k^2 \to -\infty 
\end{equation}
In 1-loop order the $\beta$-functions are calculated to be
\begin{eqnarray}
 \b _ \l 
^{ (1)} 
& = & {1 \over 8 \pi ^ 2} 
 {1 \over 3} \bigl(5 - 8 {m_{f} ^4\over m_{H} ^4}   +  4 {m_{f}^2 
\over m  _{H}^2}  \bigr) \l^2  \equiv b _\l ^{(1)} (
 \frac {\mf }{\mhi }   ) \l^2 
   \nonumber \\ 
  \b _{ m_{f}}
   ^{(1)} 
 &=&  - {1 \over 16 \pi ^ 2}  { 1\over 3}  \Bigl( 5  - 8
{m_{f} ^4\over m_{H}^4}  \Bigr) \l \equiv b_ {\mf}^{(1)}
 ( \frac {\mf }{\mhi }  ) \l 
\end{eqnarray}
For solving the CS and RG equation consistently their commutators
have to vanish which gives the following restrictions on the
$\beta$-functions: 
\begin{equation}
\k\p_\k \beta _\l = \b _{m_f} m_f \p_{\mf}\b_\l \qquad
\k\p_\k \beta _{m_f} = \b _{\l} \p_\l \b_{m_f}
\end{equation} \pagebreak[2]
Therefrom it is seen
that the $\beta $-functions logarithmically depend on the normalization point 
from 2-loop order onwards.
%\begin{equation}
%\b^{(n+1)}  
%\sim \ln ^n |\frac {m^2}{\k^2 }  | \quad \hbox{for} \quad \k^2 \to -\infty 
%\end{equation}
For the 2-loop functions  one obtains the result\cite{InvCh}:
\begin{eqnarray}
\b_{\l}  
^{(2)}  
& =& \l^3 \Bigl(
- \frac 12  b _{\mf} ^{(1)} (\hbox{$\mf \over \mhi$}) m_f \p_ {\mf}
b ^{(1)} _\l (\hbox{$\mf \over \mhi$}) \ln   | \frac {\mhi^2}{\k^2}  |
+ b ^{(2)} _\l (\hbox{$\mf \over \mhi$}) \Bigr) \nonumber \\
\b
^{(2)} _{\mf} & =  &  \l^2 \Bigl(   \frac 12 
 b _{\mf} ^{(1)} (\hbox{$\mf \over \mhi$}) 
b ^{(1)} _\l (\hbox{$\mf \over \mhi$})
\ln | \frac {\mhi^2}{\k^2}  | +  b _{\mf} ^{(2)} (\hbox{$\mf \over \mhi$})
\Bigr)
\end{eqnarray}
$b_\l^{(2)}$ and $b_{\mf}^{(2)}$ are normalization point independent
2-loop contributions.
The invariant charge can be constructed as the simultaneous solution
of the CS and RG equation, if one takes into account the mass dependent
1-loop induced contributions of the consistency equation. Inserting
in the RG invariant solution (\ref{rginv}) one gets in 1- and 2-loop
order for the  invariant charge at asymptotic $p^2$ and $\kappa ^2$:
\begin{eqnarray}
 Q ^ {(1)} _{{as}}
 & = & \frac 12 \l^2 b _ \l
^{(1)}( \hbox{$\mf \over \mhi$} )  \ln  (\hbox{$ {p^2}\over {\k^2}  $} )
\nonumber \\
Q ^{(2)}_{{as}}
 & = & \l^3  \biggl(
\Bigl(\frac 12
b ^{(1)} _\l (\hbox{$\mf \over \mhi$}) \ln (\hbox{$ {p^2} \over {\k^2}$} )
  \Bigr)^2
+ \frac  12
b ^{(2)} _\l (\hbox{$\mf \over \mhi$} ) \ln ( \hbox{$ {p^2}\over {\k^2} $} )
 \nonumber \\
&  &  - \frac 18  b _{\mf} ^{(1)} (\hbox{$\mf \over \mhi$}) \mf \p_ {\mf}
b ^{(1)} _\l (\hbox{$\mf \over \mhi$}) 
\Bigl(\ln ^2  |\hbox{$ {\mhi^2}\over {\k^2}$ }  |
- \ln ^2 | \hbox{$ {\mhi^2}\over {p^2} $} | \Bigr) \biggr)
\end{eqnarray}
This shows that the 2-loop RG invariant contains quadratic logarithms
induced by the CS equation. They appear due to the fact that the
fermion mass has to be treated as an independent parameter of the model,
 being differently renormalized
when compared to 
the coupling constant of the Higgs 4-point interaction. RG invariance
can be only formulated  for the coupling $\l$ and does not include
the fermion mass
 in order to be able to construct the S-matrix of the theory.
 The asymptotic 
behavior of the spontaneously broken model differs from the symmetric
theory  
 by these mass dependent logarithms.
As a consequence an asymptotic limit in the sense of mass independence
does not exist: The asymptotic normalization conditions are defined by
the requirement that the terms of order $m^2 / \k^2 \ln | m^2 /\k^2| $
can be neglected. The smaller these terms are chosen the larger the
mass dependent logarithms in the $\beta$-functions will grow.
Therefore the massless symmetric theory is not the asymptotic version
of the spontaneously broken one.

\section{The Callan--Symanzik equation of the standard model}
In the standard model of electroweak interactions the masses of
all massive par\-ticles are generated by the spontaneous breaking
of $SU(2) \times U(1)$ gauge invariance. The massive gauge bosons
are the
 charged bosons $W_{\pm}$ with mass $M_W$ and the neutral boson $Z$
with mass $M_Z$.
The minimal standard model contains a complex scalar doublet $\Phi$ with
three unphysical would-be Goldstones and the physical field $H$
with mass $m_H$. In addition it involves also the electromagnetic
interactions coupled to the massless photon field $A_\mu$.
In  the fermionic sector there are the
left handed lepton and quark doublets and right handed singlets. For 
simplification we do not consider mixing between different families
and, especially, assume CP-invariance.

In order to have free field propagators with a
good ultraviolet and infrared behavior one  uses
 the linear $R_\xi$-gauges 
breaking thereby $SU(2) \times U(1)$ gauge invariance
and also rigid invariance. Gauge invariance has to be replaced
by BRS invariance introducing the Faddeev--Popov ghosts
$c_a$ and $\bar c_a, a= +,-,Z,A$. 
The Green functions have to
be constructed in accordance with the Slavnov--Taylor (ST) identity, 
which is the functional version of BRS invariance. 
It turns out that the Green functions are not completely determined
by the ST identity -- in addition one has to use the Ward identity
of rigid invariance and the remaining $U(1)$-local Ward identity
in order to fix the charges of the fermions.
Using an on-shell scheme for determining the free parameters,
one has  one coupling, which can be
fixed  in the Thompson limit to be the fine structure
constant $\alpha= e^2 / 4\pi $. 
\begin{equation}
M_W, \, M_Z,\, m_H, \,  m_f, \, e
\end{equation}
are the free parameters of the standard model,
which are complemented by two gauge parameters and the masses of
the ghosts 
\begin{equation}
\xi , \hat \xi , \zeta_W M_W , \zeta_Z M_Z
\end{equation}
With this choice of parameters a normalization point $\kappa^2$
is only introduced for fixing infrared divergent residua off-shell
 and
the RG equation  is trivial
concerning the coupling.

The CS equation of the standard model gives the breaking of
dilatations in form of a differential equation  the same
way as   in the $U(1)$-axial model. 
In the tree approximation dilatations are broken by the mass terms
of the fields
They are related to the 
 differentiation with respect to the Higgs field and to the external
Higgs field
which
had to be introduced for establishing  the Ward identity of rigid symmetry:
\begin{equation}
m\partial_m \Gamma_{cl} = \int v \Bigl( {\delta \Gamma_{cl}\over \delta H } +
\zeta {\delta \Gamma_{cl}\over \delta \hat H } \Bigr) +  
\frac {m_H^2}2  \Delta_{inv} 
\end{equation}
 $\Delta_{inv}$ is the 2-dimensional BRS and rigid invariant 
scalar polynomial.
The r.h.s.~potentially contains non-integrable infrared  divergencies, 
which have to be proven to be absent. Therefore it is necessary to
have  propagators which are well-behaving in the infrared,
i.e.~one has to use the physical fields and has to impose to
all orders that the mixing between massless and massive fields
vanishes at $p^2 = 0$. In the ghost sector this requires to introduce
an independent angle $\theta_G$ defined by the mass ratio of the
ghosts:
\begin{equation}
1 + \tan \theta_ W \tan \theta _G \equiv {\zeta _Z M _Z^2  \over
\zeta _W M_W^2} \qquad \cw \equiv {M_W \over M_Z}
\end{equation} 
%\begin{eqnarray}
%\label{infra}
%\Gamma_{ZA}(p^2=0) = \Gamma_{AA}(p^2 = 0) &=& 0 \nonumber\\
%\Gamma_{\bar c _A c _Z}(p^2=0) =
%\Gamma_{\bar c _Z c _A}(p^2=0) = \Gamma_{\bar c_ Ac _A}(p^2 = 0) &=& 0 
%\end{eqnarray}

 The CS equation is uniquely determined if one
uses invariance of the Green functions with respect to the
ST identity, rigid symmetry and $U(1)$-gauge symmetry.
For simplification we neglect all the contributions which involve
Faddeev--Popov ghosts and external fields. In 1-loop  order the CS equation 
 then reads\cite{KrWe}:
\begin{eqnarray}
&\hspace{-3mm} &
 \hspace{-3mm}\biggl\{
 m \partial _{ m} +\beta_e e\p_e +
\beta _{m_H}m_H \p _{m_H}  
- \beta_{ {M_W}}
\Bigl(   \p_{\theta _W}
-   \int\! \bigl(
Z^\mu {\delta \over \delta A^\mu } 
 - A^\mu {\delta \over \delta Z^\mu } \bigr) 
\Bigr)  
\nonumber \\
&\hspace{-3mm}&\hspace{-3mm} - \hat\gamma_{V} \Bigl( \int \!(
 {\sin \theta_W}
 Z^\mu  + \cw A^\mu   )
 \bigl( \sw {\delta \over \delta Z^\mu } + \cw {\delta \over \delta A^\mu }  
\bigr)
+ 2 (\hat \xi + \xi )\p _{\hat \xi} \Bigr)\nonumber \\
&\hspace{-3mm} &\hspace{-3mm} - \gamma_V \Bigl(\N_V
 + 2 \xi (\partial _\xi +  \partial_{\hat \xi}) + \sg \cg \p_{\theta_G}
\Bigr) 
- \gamma _S \N_S \nonumber \\
&\hspace{-3mm} &\hspace{-3mm} 
+ \sum_{f_i}\bigl(\beta _{m_{f_i}} 
m_{f_i}\p _{m_{f_i}} - \gamma^R_{f_i} \N_{f_i}\bigr)    
- \sum_{F_i}\gamma^L _{F_i} \N_{F_i}
\biggr\} \Gamma \bigg|_{ext.f.= 0 \atop c_a,\bar c_a =0}
=  [\Delta_s]_3 ^3 \cdot \Gamma 
\end{eqnarray} 
There $[\Delta_s]_3 ^3 $ denotes the soft insertion, which uniquely
continues the classical soft breaking of dilatations to higher orders.
The operators $\N_A $ are the usual leg counting operators of vectors,
scalars, right handed fermions and left handed fermion doublets.
The $\beta$-functions of the electromagnetic coupling and the
of the mass ratio $M_W \over M_Z$ are calculated to
\begin{eqnarray}
\beta_e & =& -{\alpha \over 24 \pi} \Bigl( 42 - \hbox{$\frac{64}3$} N_F \Bigr)
\nonumber \\
\beta_{M _W} 
& = & - {\alpha \over 24 \sw \cw} \Bigl( (43 - 8 N_F) - (42 - 
                     \hbox{$\frac {64}3$} N_F ) \sin^2 \theta_W \Bigr)
\end{eqnarray}

One has to note that the CS equation of the standard model has
unusual ingredients compared to the symmetric theory: It involves
the mixed field differentiation operators due to the fact the unbroken
electromagnetic symmetry has also  SU(2) interactions. Especially
there appear the $\beta$-functions with respect to  the different
masses of the standard model. Because these $\beta$-functions
are mass dependent already in one loop order, the 2-loop 
$\beta$-functions  depend on mass logarithms in the same way
as we have derived in the $U(1)$-axial model.

\vspace*{0.5cm}
{\it Acknowledgment}
I am grateful to B.A.~Kniehl for
the opportunity to participate in
 this interesting workshop.

\end{document}